\documentclass[journal]{IEEEtran}

\usepackage{graphics}
\usepackage{epstopdf}
\usepackage[pdftex]{graphicx}
\usepackage{stfloats}
\usepackage{array}
\usepackage{float}
\usepackage[cmex10]{amsmath}
\usepackage{amsmath,amsfonts,amssymb}
\usepackage{graphicx,graphics}
\usepackage{enumerate}
\usepackage{color}
\usepackage{verbatim}
\usepackage{amsthm}
\usepackage{multirow}
\usepackage{subcaption}
\usepackage{siunitx}
\usepackage{amsmath,amsthm}
\usepackage{cleveref}
\usepackage{cite}
\usepackage[font=footnotesize]{caption}
\usepackage{fixltx2e}
\usepackage[USenglish]{babel}
\usepackage{lettrine}

\newcommand{\e}{\mathrm{e}}

\newcommand{\erf}{\mathrm{erf}}
\begin{document}
\title{Performance Analysis of Analog Intermittently Nonlinear Filter in the Presence of Impulsive Noise }

\author{
Reza Barazideh$^{\dag}$, Balasubramaniam Natarajan$^{\dag}$, Alexei V. Nikitin$^{\dag \ast}$, Solmaz Niknam$^{\dag}$, \\
\small $^{\dag}$ Department of Electrical and Computer Engineering, Kansas State University, Manhattan, KS, USA.\\
$^{\ast}$ Nonlinear LLC, Wamego, KS 66547, USA.\\
Email:\{rezabarazideh, bala, slmzniknam\}@ksu.edu, avn@nonlinearcorp.com}

\maketitle

\begin{abstract}

An Adaptive Nonlinear Differential Limiter (ANDL) is proposed in this paper to efficiently alleviate the impact of impulsive noise (IN) in a communication system. Unlike existing nonlinear methods, the ANDL is implemented in the analog domain where the broader acquisition bandwidth makes outliers more detectable and consequently it is easier to remove them. While the proposed ANDL behaves like a linear filter when there is no outlier, it exhibits intermittent nonlinearity in response to IN. Therefore, the structure of the matched filter in the receiver is modified to compensate the filtering effect of the ANDL in the linear regime. In this paper, we quantify the performance of the ANDL by deriving a closed-form analytical bound for the average signal-to-noise ratio (SNR) at the output of the filter. The calculation is based on the idea that the ANDL can be perceived as a time-variant linear filter whose bandwidth is modified based on the intensity of the IN. In addition, by linearizing the filter time parameter variations, we treat the ANDL as a set of linear filters where the exact operating filter at a given time depends upon the magnitude of the outliers. The theoretical average bit error rate (BER) is validated through simulations and the performance gains relative to classical methods such as blanking and clipping are quantified.

\end{abstract}

\begin{IEEEkeywords}
Impulsive noise (IN), analog nonlinear filter, adaptive nonlinear differential limiter (ANDL), orthogonal frequency-division multiplexing (OFDM).
\end{IEEEkeywords}

\section{Introduction}

\lettrine[lines=2]{M}{ulticarrier} transmission techniques have been proposed to cope with the frequency selectivity of the propagation channel in many applications \cite{MCM}. Particularly, orthogonal frequency-division multiplexing (OFDM) is widely used in many applications in vehicular communications ranging from wired communication such as Power-line communication (PLC) in Home-Plug Green PHY standard for V2G communications \cite{HomePlug} to wireless communications such as 802.11p Wireless Access in Vehicular Environments (WAVE) standard \cite{Standard_Wave}, and underwater acoustic (UWA) communication \cite{Imp_UWA_Kuai_2016}. However, OFDM provides some level of robustness against impulsivity, system performance can still degrade if the impulsive noise (IN) exceeds a certain threshold and its effect gets spread over all subcarriers \cite{ghosh1996}. Taking an OFDM-based system as an example, this paper introduces and analytically quantifies the performance of an analog intermittently nonlinear filter in the presence of IN.

\subsection{Related work}

Many techniques have been explored in prior efforts to mitigate the impact of IN. For example, robust iterative channel decoding techniques have been used to ameliorate bit error rate (BER) in impulsive environments \cite{Turbo2014Umehara, LDPC2005Nakagawa}. It has been shown that coding techniques are mostly effective in single carrier schemes and there is no gain in OFDM systems \cite{GNdo_TurboCode_2008}. In addition, frequency or time domain interleaving ~\cite{Nassar12local, Time_Interleaving, Liu_Opt_Interleaving_2016} are not effective in highly impulsive environments. Moreover, compressive sensing (CS) techniques are used to estimate IN by measurements on null subcarriers of OFDM \cite{CS_2008, CS_Alnafouri_2014,DoubleKill_Liu_2016, Liu_Structured_CS_2017}. In \cite{Lin13impulsive_SparseBayesian} a non-parametric algorithm is proposed by extension of \cite{CS_2008} to a sparse Bayesian learning (SBL) approach \cite{SBL_Wipf_2004}. A combination of factor-graph-based receiver and message-passing technique \cite{GAMP_Rangan_2011} is proposed in \cite{Factor_graph_2014} to mitigate IN.

High amplitude and short duration of IN has also motivated the use of various memoryless nonlinear approaches such as clipping \cite{Tseng-2012-robust-clipping}, blanking \cite{Blanking,Blanking_Zhidkov_2006}, linear combination of blanking and clipping \cite{LinearComBLNCLP2016Juwono}, deep clipping \cite{DeepCliping_2014_Juwono}, and multiple-threshold blanking/clipping \cite{MultiThershold_Rozic_2018}.
The output signal-to-noise ratio (SNR) can be maximized by optimizing the thresholds used in the memoryless nonlinear approaches. However, the clipping and blanking thresholds are usually experimentally derived. In \cite{AdaptiveNoiseMitigation-2010}, a threshold optimization based on Neyman-Pearson criterion is proposed and an analytical equation for the quasi-optimal blanking and clipping thresholds is provided in \cite{DesignBLNCLP_Oh_2017}.
%As shown in \cite{Zhidkovn08_Simpleanalysis}, these methods are effective only for asynchronous impulsive noise in high signal-to-impulsive noise ratios (SIR) and their performance degrades dramatically in severe impulsive environments.
%To address the challenge of severe impulsive noise conditions, a computationally intensive two-stage nulling algorithm based on iterative channel estimation is proposed in \cite{Two_Stage_Iterative}.
Bandwidth reduction in the process of analog-to-digital conversion (ADC) is the main drawback of all these digital nonlinear approaches. Therefore, in our prior works we proposed using Blind Adaptive Intermittently Nonlinear Filters (BAINFs) to mitigate the IN before the ADC. An Adaptive Nonlinear Differential Limiter (ANDL) is considered as one realization of BAINFs and the basics of ANDL are studied in \cite{Alexi-Dale}, \cite{Khodam_Latincom}. A practical implementation of BAINFs as Adaptive Canonical Differential Limiter (ACDL) along with matched filter modification is discussed in \cite{Khodam_ICC} to mitigate the IN in PLC system in real time.

\subsection{Contributions}
In this paper, a simplified blind adaptive intermittently nonlinear filter architecture is proposed and a unique approach to analyse its performance is introduced. The main contributions of this work can be summarized as follows:
\begin{itemize}

\item Introducing a proper model for IN which captures its characteristics in analog domain while maintaining equivalency with the common models used in discrete domain.

\item In order to reduce the complexity of the analytical derivations, the proposed ANDL is simplified. However, we show that this simplification does not degrade the performance of the proposed filter.

\item The BER performance of the ANDL is analytically quantified by approximating the ANDL as a set of linear filters. Here, the exact linear filter that operates at a given time depends upon the magnitude of the outliers. Then, a closed-form analytical bound is derived for the average SNR at the output of the proposed filter and the analytical BER performance is validated by simulation.
\end{itemize}

The improvement in SNR and BER is due to the fact that, unlike classical IN mitigation methods, ANDL is implemented in the analog domain where the outliers are still broadband and distinguishable. Disproportional effect of the ANDL on the signal of interest and IN increases the SNR in the desired bandwidth by reducing the spectral density of the IN without significantly affecting the desired signal. The theoretical performance of the ANDL is validated via simulation of an OFDM-based system in IN environments. Moreover, we highlight the superiority of our approach over conventional techniques such as blanking, clipping, and linear filtering.

%\subsection{Organization}
The remainder of this paper is organized as follows. Section \ref{sec:System Model} describes the system and noise models. Section III presents the fundamentals of the ANDL along with matched filter modification and resolution parameter calculation. Linear approximation of the ANDL and output SNR derivations are detailed in section IV. Section V presents theoretical and simulation results and finally conclusions are drawn in Section VI.

\section{System and Noise Models}\label{sec:System Model}

\begin{figure*}
\centering
\includegraphics[scale=.44]{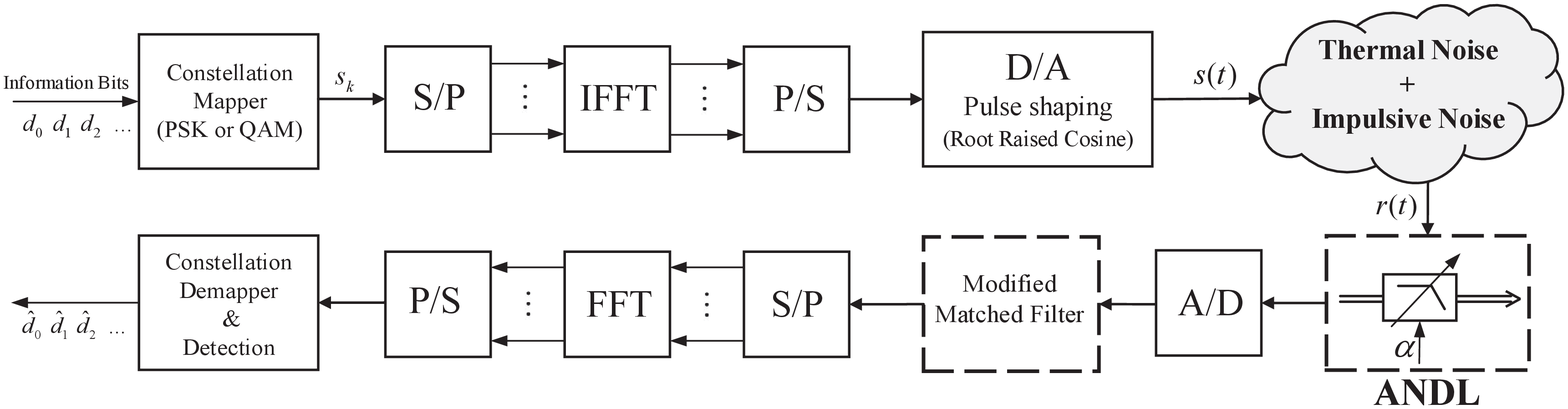}
\caption{System model block diagram.}
\label{fig:System Model}
%\vspace{.2cm}
\end{figure*}

Fig.~\ref{fig:System Model} illustrates the block diagram of the considered OFDM-based system. Here, the modulated data $s_k$ is passed through an inverse discrete Fourier transform (IDFT) to generate OFDM symbols. A root raised cosine (RRC) waveform with roll-off factor $\beta$ is used to shape and transmit the OFDM signal through the channel. The transmitted analog signal envelope in time domain can be expressed as

%-------------------------------------------------------------------
\begin{equation}
s(t) = \frac{1}{{\sqrt N }}\sum\limits_{k = 0}^{N - 1} {{s_k}}\,{\e^{\,j\frac{{2\pi kt}}{{{T}}}}}p(t),\,\,\,\,\,0 < t < {T},
\end{equation}
%-------------------------------------------------------------------
where $N$ represents the number of subcarriers; $p(t)$ denotes the pulse shape, and $T$ is the active OFDM symbol duration. Under perfect synchronization, the received signal in an additive noise channel is given by
%Without loss of generality, the power of transmitted signal is normalized to unity $\sigma _s^2 = 1$.
%In general, for different applications, we can construct an OFDM symbol with $M$ non-data subcarriers and $N-M$ data subcariers. The non-data subcarriers are either pilots for channel estimation and synchronization, or null for spectral shaping and inter-carrier interference reduction.

%-------------------------------------------------------------------
\begin{equation} \label{eq:recived signal}
r(t) = s(t) + w(t) + i(t).
\end{equation}
%-------------------------------------------------------------------
Here, $s(t)$ denotes the desired signal with variance $\sigma _s^2$ and bandwidth $B_s$; $w(t)$ is complex Gaussian noise with mean zero and variance $\sigma _w^2$; and $i(t)$ represents the IN with mean zero and variance $\sigma _i^2 \gg \sigma _w^2$. Without loss of generality, since the main objective of this paper is to demonstrate a novel approach to mitigate IN, the effect of channel fading is eliminated in \eqref{eq:recived signal}. According to the structure of the receiver in Fig.~\ref{fig:System Model}, the proposed ANDL is implemented before the ADC as a front end filter and the matched filter is modified to compensate the filtering effect of the ANDL in linear regime. In the following, we begin with a review of the impulse noise model.

\subsection{Impulsive Noise Model}

\begin{figure}[t]
\centering
\includegraphics[width=.45\textwidth,height=47mm]{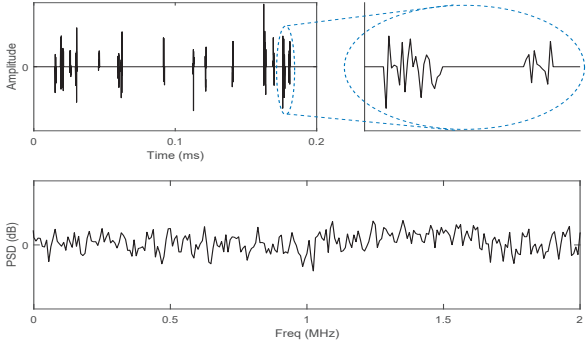}
\caption{Asynchronous impulsive noise}
\label{fig:Impulsive noise}
%\vspace{-.25cm}
\end{figure}

The widely used IN models assume the presence or absence of a strong noise component as the realization of two mutually exclusive events \cite{Zhidkovn08_Simpleanalysis }. To analyze and evaluate system performance, we propose a model that captures characteristics of an IN in the analog domain. The considered IN consists of short duration high powered impulses with random arrivals and corresponds to

%-------------------------------------------------------------------
\begin{equation}\label{eq:Impulsive_Noise}
i(t) = \nu (t)\sum\limits_{k = 1}^\infty  {{A_k}} {\kern 1pt} \left[ {\theta (t - {t_k}) - \theta (t - {t_k} - {\tau _{as}})} \right].
\end{equation}
%-------------------------------------------------------------------
Here $\nu(t)$ represents complex white Gaussian noise process with zero mean; $A_k$ is the amplitude of ${k^{th}}$ pulse and modeled by Gaussian random variable; $t_k$ is a arrival time of a Poisson process with parameter $\lambda$; $\theta(t)$ denotes the Heaviside unit step function, and $\tau_{as}$ is the duration of IN. In general the duration $\tau_{as}$ can change randomly for each burst but here, for simplicity, we assume a fixed average duration for all bursts. However, it is important to note that the method and results presented in this work can be easily extended to the case when the IN duration is random. The resulting time and frequency domains representation of this noise in analog domain is depicted in Fig.~\ref{fig:Impulsive noise}.

Note that, while \eqref{eq:Impulsive_Noise} captures a bursty IN with random amplitude in analog domain, it also can represent Bernouli-Gaussian IN model in time duration $T$ with average success probability $\varepsilon$ given by
%-------------------------------------------------------------------
\begin{align}\label{eq:Epsilon} \notag
{\varepsilon} &= {{\left[ {\sum\limits_{k = 0}^\infty  {\frac{{{{\rm{e}}^{ - \lambda T}}{{(\lambda T)}^k}}}{{k!}}k{\tau _{as}}} } \right]} \mathord{\left/
 {\vphantom {{\left[ {\sum\limits_{k = 0}^\infty  {\frac{{{{\rm{e}}^{ - \lambda T}}{{(\lambda T)}^k}}}{{k!}}\,k\,{\tau_{\rm{as}}}} } \right]} T}} \right.
 \kern-\nulldelimiterspace} T} \\\notag
&= \lambda \tau_{\rm{as}}\left[ {\sum\limits_{k = 1}^\infty  {\frac{{{{\rm{e}}^{ - \lambda T}}{{(\lambda T)}^{k - 1}}}}{{\left( {k - 1} \right)!}}} } \right]\\\notag
&= \lambda \tau_{\rm{as}}\left[ {\sum\limits_{k = 0}^\infty  {\frac{{{{\rm{e}}^{ - \lambda T}}{{(\lambda T)}^k}}}{{k!}}} } \right]\\
&= \lambda \tau_{\rm{as}}.
\end{align}
%-------------------------------------------------------------------

 In the next section, we discuss the design and implementation of ANDL in detail.
%-------------------------------------------------------------------
%-------------------------------------------------------------------

\section{Fundamentals of ANDL}\label{sec: NDL Basics}

An introduction to the fundamentals of the ANDL and finding an efficient value for the resolution parameter is provided in this section.

\subsection{ANDL Design}

%As discussed in our previous work \cite{Alexi-Dale,nikitin2014method}, the ANDL can be implemented as a first order adaptive analog nonlinear filter with time varying $\tau(t)$ which should be tuned proportionally to the amplitude of the impulsive noise. Thus, we have

\begin{figure}[t]
\centering
\includegraphics[width=.5\textwidth,height=40mm]{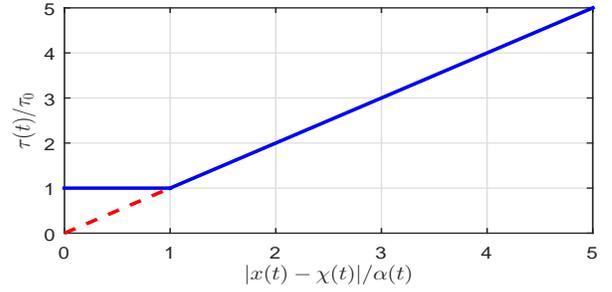}
\caption{ANDL time parameter ${\tau(t)= \tau(|x(t)-\chi(t)|)}$.}
\label{fig:Tau_Orginal}
%\vspace{-.3cm}
\end{figure}

\begin{figure}[t]
\centering
\includegraphics[width=.5\textwidth,height=60mm]{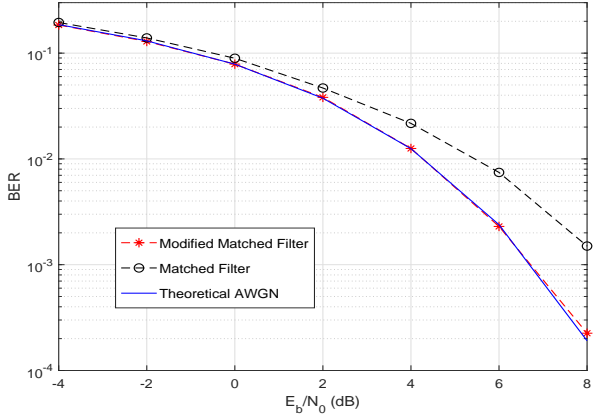}
\caption{Performance comparison between matched filter and modified matched filter in the presence of ANDL for BPSK modulation. $\beta \,{=}\, 0.25$, $\tau_0\,{=}\,1/(4\pi B_s)$.}
\label{fig:MMFvsMF}
%\vspace{-.25cm}
\end{figure}

ANDL is a blind adaptive intermittently nonlinear filter that, can be perceived as a first order time varying linear filter. According to the basic concept of the proposed ANDL \cite{Alexi-Dale,Khodam_Latincom}, the time parameter $\tau(t)$ varies proportionally with the magnitude of the difference between input and output of the filter. Therefore, we have
%-------------------------------------------------------------------
\begin{equation} \label{eq:1st order CDL}
  \chi(t) = x(t) - \tau(|x(t)-\chi(t)|)\, \dot{\chi}(t)\,,
\end{equation}
%-------------------------------------------------------------------
 where $x(t)$ and $\chi(t)$ are the input and output of the filter, respectively, and $\dot{\chi}(t)$ denotes the first time derivative of $\chi(t)$. As shown in Fig.~\ref{fig:Tau_Orginal}, the time parameter~${\tau(t)= \tau(|x(t)-\chi(t)|)}$ is given by
%-------------------------------------------------------------------
\begin{equation} \label{eq:CDL tau}
  \tau(|x(t)-\chi(t)|)  = \tau_0 \times \left\{
  \begin{array}{cc}
    \!\! 1 &|x(t)-\chi(t)| \le \alpha(t)\\
    \!\!\frac{|x(t)-\chi(t)|}{\alpha(t)} & \mbox{otherwise}
  \end{array}\right.,
\end{equation}
%-------------------------------------------------------------------
where $\tau_0$ is a fixed time constant and $\alpha(t)$ is the resolution parameter of the filter. The value of $\alpha(t)$ should be determined properly in order to mitigate the IN efficiently. In general, the ANDL is an intermittent nonlinear filter and behaves linearly, when the magnitude of the difference signal $\left| {x(t) - \chi(t)} \right|$ remains within a certain range determined by the resolution parameter $\alpha(t)$. This allows us to avoid instabilities that are often associated with nonlinear filtering. However, in case of outliers, the proper selection of $\alpha(t)$ leads the ANDL to the nonlinear regime to suppress the outliers. Based on \eqref{eq:CDL tau}, ANDL is extremely aggressive toward high amplitude IN, i.e., larger spikes in the input signal will result in a greater suppression at the output.
%Therefore, proper selection of resolution parameter can ensure that when the outliers are encountered, the nonlinear response of the ANDL suppresses the magnitude of the respective outliers in the output signal.
%--------------------------------------------------------------------------
\noindent
%--------------------------------------------------------------------------
It is worth noting that, we extend the works in \cite{Alexi-Dale,Khodam_Latincom} by adding matched filter modification module to compensate for the ANDL in the linear regime. The impulse response $h_{\rm mod}[k]$ of the modified matched filter in the discrete domain can be expressed as \cite{Khodam_ICC},
%-------------------------------------------------------------------
\begin{equation}\label{eq:MMF}
h_{{\rm mod}}[k] = h[k] + \tau_0 \dot h[k],
\end{equation}
%-------------------------------------------------------------------
where $h[k]$ is the impulse response of the matched filter and $\dot h[k]$ denotes the first time derivative of $h[k]$. The modification is done in the discrete domain, as this reduces the computational complexity and neglects the need for extra hardware components. The compensation of the modified matched filter on the BER performance of an OFDM system with $B_s=100 kHz$ and binary phase shift keying (BPSK) modulation is shown in Fig.~\ref{fig:MMFvsMF}. A root-raised-cosine filter with roll-off factor $1/4$ is considered as matched filter in Fig.~\ref{fig:MMFvsMF}. Therefore, the performance loss of ANDL in linear regime is compensated by modified matched filter when there is no IN.
%-------------------------------------------------------------------
%-------------------------------------------------------------------

\subsection{Resolution Parameter Calculation}

According to the structure of ANDL, the objective is to determine a time-dependent resolution parameter $\alpha(t)$ that enhances the quality of non-stationary signals under time-varying noise conditions. Therefore, an efficient value of $\alpha(t)$ should allow to maximize the suppression of the IN without distorting the signal of interest. It is assumed that the power of thermal noise is fixed over one OFDM symbol duration. Therefore, the resolution parameter is constant ($\alpha(t){=}\alpha$) in the duration of each OFDM symbol and it only changes across symbols. A proper value of resolution parameter $\alpha$ can be found based on difference signal $\left| {x(t) - \chi(t)} \right|$ when there is no IN. An estimate of the aforementioned difference signal can be obtained by passing signal $s(t)+w(t)$ through a linear highpass filter. Let $z(t)$ be given by a differential equation for the first order highpass filter with the time constant~$\tau_0$. Then, we have \cite{Khodam_Latincom}
%---------------------------------------------------------
\begin{equation} \label{eq:1st high pass}
  z(t)=\tau_0 \left[ \dot{s}(t)+\dot{w}(t)-\dot{z}(t) \right].
\end{equation}
%---------------------------------------------------------
As derived in our preliminary work \cite{Khodam_Latincom}, an efficient value of the resolution parameter ${\alpha_{\rm eff,\zeta}}$ for $(1-\zeta)$ level distortionless filtering of the transmitted OFDM signal in thermal noise is given by
%---------------------------------------------------------
\begin{equation} \label{eq:alpha}
\alpha_{\rm eff,\zeta} \ge \erf^{-1}(1-\zeta)\sqrt{2} \sigma_z,
\end{equation}
%-------------------------------------------------------------------------
where $\sigma _z^2$ is the variance of $z(t)$; $\erf(.)$ represents the error function; and $\zeta$ is a sufficiently small constant (e.g., ${\zeta=4.68\times 10^{-3}}$). Now that we have summarized the structure and operation of the ANDL, in the next section we derive analytical expressions for the average SNR at the ANDL output.

\section{Linear Approximation of The ANDL }\label{sec: NDL Basics}

In order to characterize the theoretical performance of the ANDL we employ a linear approximation.

\subsection{Time Parameter $\tau(t)$ Approximation}

\begin{figure}[t]
\centering
\includegraphics[width=.5\textwidth,height=67mm]{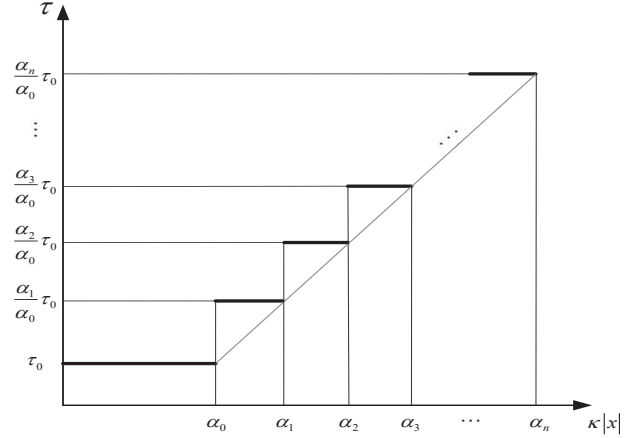}
\caption{ANDL time parameter ${\tau= \tau(\kappa|x|)}$.}
\label{fig:Tau}
%\vspace{-.3cm}
\end{figure}

\begin{figure}[t]
\centering
\includegraphics[width=.5\textwidth,height=60mm]{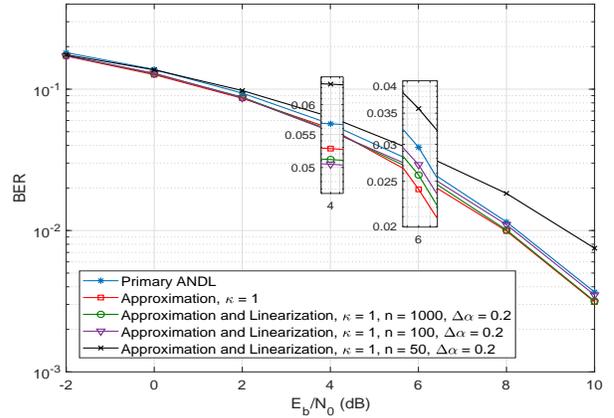}
\caption{Linear approximation of ANDL. SIR = 0 dB, $\lambda = 2B_s$, $\tau_{as}=1\mu {\rm{s}}$.}
\label{fig:Linear_Approx}
%\vspace{-.3cm}
\end{figure}

According to \eqref{eq:CDL tau}, the proposed ANDL enters the nonlinear regime only at the time of incoming IN where the difference signal $\left| {x(t) - \chi(t)} \right|$ would be approximately equal to $|x(t)|$. Therefore, the time parameter of the ANDL in \eqref{eq:CDL tau} can be approximated as
%(i.e., amplitude of impulsive noise is much greater than signal and thermal noise)
%-------------------------------------------------------------------
\begin{equation} \label{eq:CDL tau Approx}
  \tau(\kappa|x(t)|)  = \tau_0 \times \left\{
  \begin{array}{cc}
    \!\! 1 & \mbox{for} \quad \kappa|x(t)| \le \alpha_0\\
    \!\!\frac{\kappa|x(t)|}{\alpha_0} & \mbox{otherwise}
  \end{array}\right.,
\end{equation}
%-------------------------------------------------------------------
%In the simulation result it is shown that even with $\kappa=1$ the performance is satisfying.
where $\alpha_0=\alpha_{\rm eff,\zeta}$, and $\kappa$ is a positive constant that can be used to tune the modified ANDL for various IN models. In order to find the theoretical performance we approximate the ANDL by combination of $n$ linear filters as illustrated in Fig.~\ref{fig:Tau}. Here, the time constant of each individual linear filter can be expressed as
%-------------------------------------------------------------------
\begin{equation} \label{eq:Tau_Piecewise}
\tau(t)  = \left\{ \begin{array}{l}
{\tau _0},\,\,\,\,\,\,\,\,\,\,\,\,\,\,\,\,\,\,\,\,\,\,\,\,\,\kappa |x(t)|\, < {\alpha _0}\\
{\tau _1} = \frac{{{\alpha _1}}}{{{\alpha _0}}}{\tau _0},\,\,\,\,\,{\alpha _0} < \kappa |x(t)|\, < {\alpha _1}\\
 \vdots \\
{\tau _k} = \frac{{{\alpha _k}}}{{{\alpha _0}}}{\tau _0},\,\,\,\,\,{\alpha _{k - 1}} < \kappa |x(t)|\, < {\alpha _k}
\end{array} \right..
\end{equation}
%-------------------------------------------------------------------
As can be seen in Fig.~\ref{fig:Linear_Approx}, the performance of the approximated ANDL in \eqref{eq:CDL tau Approx} with $\kappa=1$ is almost the same as the primary ANDL in \eqref{eq:CDL tau}. Fig.~\ref{fig:Linear_Approx} also shows that the approximation with a combination of $n$ linear filters results in performance equivalent to the \eqref{eq:CDL tau}. Theoretically, we have the best approximation when $n\rightarrow\infty$ where the difference between two consecutive filters $\triangle\alpha=\alpha_k-\alpha_{k-1}$, $1 \leq k \leq n$ is small and the values of $\alpha_k$ are optimized. In this work, for simplicity, the linearization is performed assuming a constant $\triangle\alpha$. Fig.~\ref{fig:Linear_Approx} shows that in practice, a reasonable value of $n$ and $\triangle\alpha$ that guarantee $\alpha_n = \alpha_0+n\triangle\alpha > max|x(t)|$ (cover the entire range of $|x(t)|$) ensures the accuracy of the approximation.
%--------------------------------------------------------------------------
%--------------------------------------------------------------------------
\noindent
%--------------------------------------------------------------------------

%--------------------------------------------------------------------------

In our ANDL structure, the received signal passes through a broadband lowpass filter to limit the input noise power while ensuring that the IN is not excessively spread out in time. Considering a sufficiently broadband front end filter, the input signal $x(t)$ for ANDL can be represented by a stationary mixture of two Gaussian components weighted by $1-\varepsilon$ and $\varepsilon$. Therefore, the probability density function (PDF) of the input signal $x(t)$ can be expressed via a Gaussian Mixture (GM) model given by %according to \eqref{eq:Epsilon}
\begin{equation} \label{pdf}
{f_X}(x) = (1 - \varepsilon ){\phi _{{x_1}}}(0,\sigma _1^2) + \varepsilon {\phi _{{x_2}}}(0,\sigma _2^2),
\end{equation}
%-------------------------------------------------------------------
where
\begin{equation} \label{Weighted_Recieved Signal}
\begin{array}{l}
{x_1}(t) = s(t) + w(t)\thicksim \mathcal{N}(0,\sigma _1^2=\sigma _s^2 + \sigma _w^2) \\
{x_2}(t) = s(t) + w(t) + i(t)\thicksim \mathcal{N}(0,\sigma _2^2=\sigma _s^2 + \sigma _w^2 + \sigma _i^2),
\end{array}
\end{equation}
%-------------------------------------------------------------------
%with
%\begin{equation}
%\begin{array}{l}
%\sigma _1^2 = \sigma _s^2 + \sigma _w^2\\
%\sigma _2^2 = \sigma _s^2 + \sigma _w^2 + \sigma _i^2,
%\end{array}
%\end{equation}
%-------------------------------------------------------------------
and ${\phi _x}(.)$ is the Gaussian PDF defined by
%-------------------------------------------------------------------
\begin{equation} \label{Gaussian}
{\phi _x}(\mu ,{\sigma ^2}) = \frac{1}{{\sqrt {2\pi } \sigma }}{{\rm{e}}^{ - \frac{{{{(x - \mu )}^2}}}{{2{\sigma ^2}}}}}.
\end{equation}
%-------------------------------------------------------------------
Based on the GM model and according to \eqref{eq:Tau_Piecewise}, the average filtering effect of the ANDL can be computed via an averaged time parameter $\tau$ corresponding to

%-------------------------------------------------------------------
\begin{equation}\label{eq:Tau_Average}
\mathbb{E}[\tau]  = (1 - \varepsilon )\sum\limits_{k = 0}^n {{p_{k,1}}} {\tau _k} + \varepsilon \sum\limits_{k = 0}^n {{p_{k,2}}} {\tau _k},
\end{equation}
%-------------------------------------------------------------------
where,
%-------------------------------------------------------------------
\begin{align} \label{eq:P1}\notag
{p_{k,1}} &= \left\{ {\begin{array}{*{20}{l}}
{\Pr (0 < \kappa \left| {{x_1}(t)} \right| < {\alpha _0}),\quad \quad \,\,\,\, k = 0}\\
{\Pr ({\alpha _{k - 1}} < \kappa \left| {{x_1}(t)} \right| < {\alpha _k}), \quad k = 1,...,n}
\end{array}} \right.\\
 &= \left\{ {\begin{array}{*{20}{l}}
{1 - {\rm{erfc}}\left( {\frac{{{\alpha _0}}}{{\sqrt 2 \kappa {\sigma _1}}}} \right),\quad \quad \quad \quad \quad \,k = 0}\\
{{\rm{erfc}}\left( {\frac{{{\alpha _{k - 1}}}}{{\sqrt 2 \kappa {\sigma _1}}}} \right) - {\rm{erfc}}\left( {\frac{{{\alpha _k}}}{{\sqrt 2 \kappa {\sigma _1}}}} \right),k = 1,...,n}
\end{array}} \right.,
\end{align}

%\begin{align} \label{eq:P1}\notag
%{p_{k,1}} &= \Pr ({\alpha _{k - 1}} < \kappa \left| x_1(t) \right| < {\alpha _k}) \\ \notag
%&= 2\Pr ({\alpha _{k - 1}} < \kappa {x_1(t)} < {\alpha _k}) \\
%&= erfc\left( {\frac{{{\alpha _{k - 1}}}}{{\sqrt 2 {\kappa\sigma _{{1}}}}}} \right) - erfc\left( {\frac{{{\alpha _k}}}{{\sqrt 2 {\kappa\sigma _{{1}}}}}} \right)
%\end{align}
%-------------------------------------------------------------------
and
%-------------------------------------------------------------------
\begin{align} \label{eq:P2}\notag
{p_{k,2}} &= \left\{ {\begin{array}{*{20}{l}}
{\Pr (0 < \kappa \left| {{x_2}(t)} \right| < {\alpha _0}),\quad \quad \,\,\,\, k = 0}\\
{\Pr ({\alpha _{k - 1}} < \kappa \left| {{x_2}(t)} \right| < {\alpha _k}), \quad k = 1,...,n}
\end{array}} \right.\\
 &= \left\{ {\begin{array}{*{20}{l}}
{1 - {\rm{erfc}}\left( {\frac{{{\alpha _0}}}{{\sqrt 2 \kappa {\sigma _2}}}} \right),\quad \quad \quad \quad \quad \,k = 0}\\
{{\rm{erfc}}\left( {\frac{{{\alpha _{k - 1}}}}{{\sqrt 2 \kappa {\sigma _2}}}} \right) - {\rm{erfc}}\left( {\frac{{{\alpha _k}}}{{\sqrt 2 \kappa {\sigma _2}}}} \right),k = 1,...,n}
\end{array}} \right..
\end{align}
%\begin{align} \label{eq:P2}\notag
%{p_{k,2}} &= \Pr ({\alpha _{k - 1}} < \kappa \left| x_2(t) \right| < {\alpha _k}) \\ \notag
%&= 2\Pr ({\alpha _{k - 1}} < \kappa {x_2(t)} < {\alpha _k}) \\
%&= erfc\left( {\frac{{{\alpha _{k - 1}}}}{{\sqrt 2 {\kappa\sigma _{{2}}}}}} \right) - erfc\left( {\frac{{{\alpha _k}}}{{\sqrt 2 {\kappa\sigma _{{2}}}}}} \right)
%\end{align}
%-------------------------------------------------------------------
Here, erfc(.) represents the complementary error function.

%is given by
%\begin{equation}
%%\rm{erfc}(u) = \frac{2}{{\sqrt \pi  }}\int\limits_u^\infty  {e^{ - {t^2}}} dt.
%{\rm{erfc}}(u){\rm{ = }}\frac{{\rm{2}}}{{\sqrt \pi  }}\int\limits_u^\infty  {{{\rm{e}}^{{\rm{ - }}{t^{\rm{2}}}}}} dt.
%\end{equation}

\subsection{Output of the ANDL}
Considering \eqref{eq:Tau_Piecewise}, the ANDL can be approximated by a weighted combination of $n$ linear filters with each of them functioning with probabilities corresponding to \eqref{eq:P1} and \eqref{eq:P2}. Thus, the average output of the filter based on a mixture model input can be expressed as
%-------------------------------------------------------------------
\begin{equation}\label{eq:Filter_Output}
\chi (t) = \left\{ \begin{array}{l}
{\chi _1}(t),\,\,\,\,\rm{with \ probability}\,\,\,1 - \varepsilon \\
{\chi _2}(t),\,\,\,\,\rm{with \ probability}\,\,\, \varepsilon
\end{array}\right.,
\end{equation}
%-------------------------------------------------------------------
%Here,
%\begin{equation}\label{eq:Filter_Output2}
%\begin{array}{l}
%{\chi_1}(t) = {s_{out,1}}(t) + {w_{out,1}}(t)\\
%{\chi_2}(t) = {s_{out,2}}(t) + {w_{out,2}}(t) + {i_{out,2}}(t),
%\end{array}
%\end{equation}
%-------------------------------------------------------------------
where
\begin{equation}
\begin{array}{l}
{\chi _1}(t) = \sum\limits_{k = 0}^n {{p_{k,1}}\left\{ {\left[ {s(t) + w(t)} \right]*{h_k}(t)} \right\}},\\
{\chi _2}(t) = \sum\limits_{k = 0}^n {{p_{k,2}}\left\{ {\left[ {s(t) + w(t) + i(t)} \right]*{h_k}(t)} \right\}}.
\end{array}
\end{equation}

\begin{figure}[t]
\centering
\includegraphics[width=.38\textwidth,height=42mm]{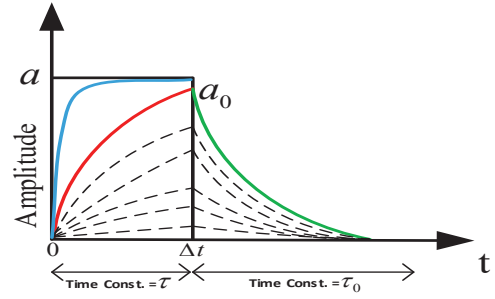}
\caption{Step Response of the ANDL.}
\label{fig:StepResponse}
%\vspace{-.3cm}
\end{figure}

%-------------------------------------------------------------------
Here, $h_k(t)$ is a first order linear lowpass filter with time constant $\tau_k$. In order to quantify the output power of each individual filter, we consider square pulses as an input (if not, each shape can be approximated by summation of narrower square pulses). According to Fig.~\ref{fig:StepResponse}, the output of the proposed ANDL consists of two parts $y_1(t)$ (red line) and $y_2(t)$ (green line) which are given by

%--------------------------------------------------------------------------
\noindent
%--------------------------------------------------------------------------
%-------------------------------------------------------------------
\begin{equation}\label{eq:StepResponse}
\begin{array}{l}
{y_1}(t)\left| {_{(\tau ,{a})}} \right.= {a} (1 - {e^{ - \frac{t}{\tau }}}),\,\,\,\,\,\,\,\,0 \le t \le \Delta t \\
\\
{y_2}(t)\left| {_{(\tau_0 ,{a})}} \right.= {a_0} {e^{ - \frac{(t-\Delta t)}{\tau_0 }}},\,\,\,\,\,\,\,\,\,\,\,\,\,\,\,\,\,t \geq \Delta t ,
\end{array}
\end{equation}
%-------------------------------------------------------------------
where $\tau$ is the time parameter for $y_1(t)$ (i.e., $\tau_k$ in $k^{\rm{th}}$ region of \eqref{eq:Tau_Piecewise}); $\tau_0$ represents the time constant and it is determined based on the bandwidth of desired signal; $\Delta t$ is duration of square pulse with amplitude $a$, and $a_0=a(1 - {e^{ - \frac{\Delta t}{\tau }}})$. Note that $\tau=\tau_0$ when there is no IN. Thus, given $\tau$, $\tau_0$ and $a$, the corresponding output power after lowpass filtering for a single pulse is given by
%-------------------------------------------------------------------
\begin{align} \label{eq:Output_Power}\notag
P\left| {_{(\tau ,{a})}} \right.&= \left( {{P_1} + {P_2}} \right)\left| {_{(\tau ,{a})}} \right.= {\int\limits_0^{\Delta t} {\left| {{y_1}} \right|} ^2}dt + {\int\limits_{\Delta t}^\infty  {\left| {{y_2}} \right|} ^2}dt\\ \notag
&= {\int\limits_0^{\Delta t} {\left| {{a}(1 - {e^{ - \frac{t}{\tau }}})} \right|} ^2}dt + {\int\limits_{\Delta t}^\infty  {\left| {{a_0}{e^{ - \frac{(t-\Delta t)}{\tau_0 }}}} \right|} ^2}dt\\ \notag
&= a^2\left[ {\Delta t - \frac{\tau }{2}{e^{ - \frac{{2\Delta t}}{\tau }}} + 2\tau {e^{ - \frac{{\Delta t}}{\tau }}} - 3\frac{\tau }{2}} \right]+ a_0^2 \frac{\tau_0 }{2}.
%&\hspace{.5cm}+ a_0^2 \frac{\tau_0 }{2}.
\end{align}
%-------------------------------------------------------------------
This amount of power is the total residual power after filtering which consists of power of the desired signal, thermal, and impulsive noises. In order to find their individual contributions, we use average residual power for desired signal and thermal noise but for IN we calculate the residual power for each region in Fig.~\ref{fig:Tau}, separately. Since the ANDL is approximated by a set of linear filters and the amplitude variation of the desired signal is much smaller than IN variation (lower bandwidth), the average residual power of desired signal can be determined by averaging over $\tau$ and $a$, that is
%averaging out over the of averaged $|s(t)|$ over all possible filters, that is
%-------------------------------------------------------------------
\begin{equation}
{P_s} = {\mathbb{E}_{\tau ,a}}[P\left| {_{\tau ,a}} \right.] =\int\int{{P\left| {_{\left( {\tau ,a} \right)}} \right..{f_T}(\tau ).{f_A}(a)\,d\tau da}}.
\end{equation}
%-------------------------------------------------------------------
In the case of the desired signal, random variable $a$ corresponds to $|s(t)|$ which has a folded-normal distribution ($s(t)$ has Gaussian distribution). Therefore, we have
\begin{align} \label{eq:Signal_Output_Power}\notag
{P_s} &={\mathbb{E} ^2}\left[ {\left| s(t) \right|} \right]\Bigg( (1 - \varepsilon )\sum\limits_{k = 0}^n {p_{k,1}}P\left| {_{{(\tau _k,1)}}} \right. \\
&\hspace{3cm}+ \varepsilon \sum\limits_{k = 0}^n {{p_{k,2}}P\left| {_{{(\tau _k,1)}}} \right.}   \Bigg),
\end{align}
%-------------------------------------------------------------------
%\begin{align} \label{eq:Signal_Output_Power}
%{P_s} &={\mathbb{E} ^2}\left[ {\left| s(t) \right|} \right]\Bigg( (1 - \varepsilon )\sum\limits_{k = 0}^n {p_{k,1}}P\left| {_{{(\tau _k,1)}}} \right. + \varepsilon \sum\limits_{k = 0}^n {{p_{k,2}}P\left| {_{{(\tau _k,1)}}} \right.}   \Bigg),
%\end{align}
%-------------------------------------------------------------------
where
\begin{equation}
\mathbb{E}[|s(t)|]= {\sigma _s}\sqrt {\frac{2}{\pi }} {e^{( - \mu _s^2/2{\sigma_s ^2})}} + {\mu _s}(1 - 2\phi (\frac{{ - \mu_s }}{\sigma_s})).
\end{equation}
Similarly, in the case of thermal noise, the random variable $a$ corresponds to $|w(t)|$ and we have
%-------------------------------------------------------------------
\begin{align} \label{eq:Noise_Output_Power}\notag
{P_w} &={\mathbb{E} ^2}\left[ {\left| w(t) \right|} \right]\Bigg( (1 - \varepsilon )\sum\limits_{k = 0}^n {p_{k,1}}P\left| {_{{(\tau _k,1)}}} \right. \\
&\hspace{3cm}+ \varepsilon \sum\limits_{k = 0}^n {{p_{k,2}}P\left| {_{{(\tau _k,1)}}} \right.}   \Bigg),
\end{align}
%%-------------------------------------------------------------------
%\begin{align} \label{eq:Noise_Output_Power}
%{P_w} &={\mathbb{E} ^2}\left[ {\left| w(t) \right|} \right]\Bigg( (1 - \varepsilon )\sum\limits_{k = 0}^n {p_{k,1}}P\left| {_{{(\tau _k,1)}}} \right. + \varepsilon \sum\limits_{k = 0}^n {{p_{k,2}}P\left| {_{{(\tau _k,1)}}} \right.}   \Bigg),
%\end{align}
%%-------------------------------------------------------------------
where
\begin{equation}
\mathbb{E}[|w(t)|]= {\sigma _w}\sqrt {\frac{2}{\pi }} {e^{( - \mu _w^2/2{\sigma_w ^2})}} + {\mu _w}(1 - 2\phi (\frac{{ - \mu_w }}{\sigma_w})).
\end{equation}
%-------------------------------------------------------------------
The amplitude variation of the IN is much larger than the amplitude variation of the desired signal and thermal noise. However, it is possible that some IN may be buried within the desired signal and thermal noise. If that is the case, then there will be no way to distinguish between IN and other components of the received signal in a band limited system. This problem highlights the advantage of the proposed ANDL which is implemented in analog domain where a wide acquisition bandwidth makes the IN more distinguishable. Thus, the absolute value of IN is more likely to be larger than the resolution parameter. Consequently, the IN will encounter a filter with large $\tau$ proportional to its amplitude as shown by dashed lines in Fig.~\ref{fig:StepResponse}. Therefore, we find the average amplitude of IN in each region of Fig.~\ref{fig:Tau} and for simplicity we pick the center of each region except in the first region where $\alpha_0$ is picked as a representative of the amplitude of IN. Thus, we have
%-------------------------------------------------------------------
\begin{align}
\mathbb{E}[|i_k|] = \left\{ {\begin{array}{*{20}{l}}
{\alpha _0,\quad \quad \quad \quad \quad \,\,\,\,\,\,\, k = 0}\\
{{\alpha _0} + \frac{{\left( {2k - 1} \right)\Delta \alpha }}{2}, \quad k = 1,...,n}
\end{array}} \right.,
\end{align}
%%-------------------------------------------------------------------
%\begin{equation}\label{eq:Impulsive_Amplitude}
%\mathbb{E}[|i_k|] = \left[ {{\alpha _0},\,\,{\alpha _0} + \frac{{\Delta \alpha }}{2},\,\,{\alpha _0} + \frac{{3\Delta \alpha }}{2},\,\,{\alpha _0} + \frac{{5\Delta \alpha }}{2}, \cdots } \right],
%\end{equation}
%-------------------------------------------------------------------
and the average residual power of IN after the linearized ANDL is given by
%-------------------------------------------------------------------
\begin{equation}\label{eq:Impulsive_Output_Power}
{P_i} = \varepsilon \sum\limits_{k = 0}^n {{\mathbb{E}^2[|i_k|]}.{p_{k,2}}.P\left| {_{{(\tau _k,1)}}} \right.}.
\end{equation}
%-------------------------------------------------------------------
Finally, the average output SNR can be expressed as
%-------------------------------------------------------------------
\begin{equation}\label{eq:Impulsive_Output_Power}
{\rm{{SNR}}}_{\rm{{avg}}} = \frac{{{P_s}}}{{{P_w} + {P_i}}}.
\end{equation}
%-------------------------------------------------------------------
Therefore, the average BER can be bounded using Jensen's inequality. For example, for BPSK ${\rm{{BER}}}_{\rm{{avg}}} \leq Q(\sqrt{2\,{\rm{{SNR}}}_{\rm{{avg}}}}) $ where $Q(.)$ is the $Q$-function.
%as a function of $SNR_{\rm{{avg}}}$ and according to Jensen's inequality the $BER_{\rm{avg}}$ is upper bounded by a concave function of $SNR_{\rm{{avg}}}$ (e.g., ).

%\begin{align}\notag
%BER_{avg} &= \mathbb{E}\left[ {f(SNR)} \right]\\
% &\leq f\left( {\mathbb{E}\left[ {SNR} \right]} \right) \leq f(SN{R_{avg}}).
%\end{align}
%%-------------------------------------------------------------------
%Here, concave function $f(.)$ depends on the modulation scheme used (e.g., for BPSK $f(SNR)=Q(\sqrt{2\,SNR}) $ where $Q(.)$ is the $Q$-function).
%%-------------------------------------------------------------------
%\begin{equation}\label{eq:Q-function}
%Q(x) = \frac{1}{{\sqrt {2\pi } }}\int\limits_x^\infty  {\exp \left( { - \frac{{{u^2}}}{2}} \right)}\, du
%\end{equation}
%%-------------------------------------------------------------------

\section{Simulation results}

In this section, the analytical results derived in the previous sections are validated through simulations. In addition, SNR and BER of an OFDM system with BPSK modulation are used to compare the performance of the proposed analog nonlinear filter to other conventional approaches such as linear filtering, blanking and clipping. As a specific example, an OFDM-based system with signal bandwidth $B_s=100\, kHz$ and $N=512$ subcarriers is chosen as a reference, but the conclusions can be extended to any OFDM system as long as the number of subcarriers is large enough to satisfy the Gaussian signal assumption. The system is investigated in an additive noise environment that consists of two components: (i) thermal noise, (ii) asynchronous random IN with normally distributed amplitudes captured by a Poisson arrival process with parameter $\lambda$ and time duration $\tau_{as}$. To mitigate the IN, a first order ANDL with ${\tau _0} {=} 1/(4\pi {B_s})$ is used. It is important to note that when $\alpha  \to \infty $ the ANDL becomes a first order linear lowpass filter and a modified matched filter is used to alleviate the filtering effect of ANDL in the linear regime. To emulate the analog signals in the simulation, the digitization rate is chosen to be significantly higher (by about two orders of magnitude) than the ADC sampling rate. Note that in all simulations, (i) the optimum thresholds for blanking and clipping are found based on an exhaustive numerical search, (ii) the resolution parameter $\alpha(t)$ for ANDL is determined based on expression \eqref{eq:alpha} with low computational complexity, and (iii) $\kappa=1$, $\Delta \alpha=0.2$, and the number of quantization levels $n$ is determined according to the dynamic range of incoming signal and considered $\Delta \alpha$.
%-------------------------------------------------------------------
\begin{figure*}
\centering
\includegraphics[scale=.53]{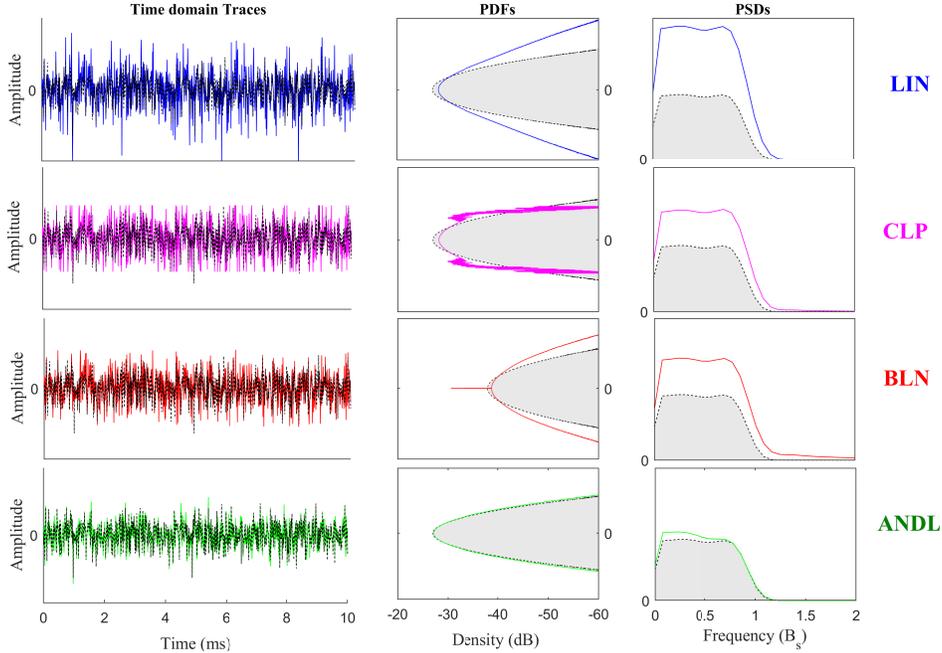}
\caption{Comparison of different approaches in time and frequency domain. $E_b/N_0$ = 10 dB, SIR = 0 dB, $\lambda=B_s$.}
\label{fig:PSD}
%\vspace{-.2cm}
\end{figure*}
%-------------------------------------------------------------------
Fig.~\ref{fig:PSD} shows the properties of the signal in time and frequency domain, and its amplitude distribution for different methods of IN mitigation. In Fig.~\ref{fig:PSD}, the black dashed lines (shaded area) represent the desired signal (without noise), and the colored solid lines represent the signal+noise mixtures. The leftmost panels show the time domain traces, the rightmost panels show the power spectral density (PSDs), and the middle panels show the amplitude densities (PDFs). From the panels of the last row, it is clear that the ANDL efficiently reduces the spectral density of the IN in the signal passband without significantly affecting the signal of interest. By comparing the panels of row LIN (Linear), CLP (Clipping) and BLN (Blanking) with row ANDL (specially PSDs panels), it can be seen that the achieved improvement due to ANDL in the quality of the baseband signal is significant. In the following, the aforementioned improvement is shown in terms of SNR and BER.
%-------------------------------------------------------------------
\begin{figure}%[t]
\centering
\includegraphics[width=.5\textwidth,height=65mm]{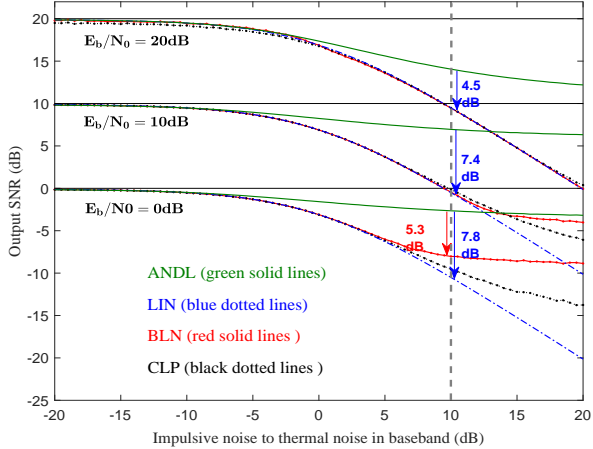}
\caption{Comparison of output SNR for different approaches. $\lambda=2B_s$.}
\label{fig:SNR}
%\vspace{-.25cm}
\end{figure}
%-------------------------------------------------------------------
%-------------------------------------------------------------------
\begin{figure}%[t]
\centering
\includegraphics[width=.5\textwidth,height=65mm]{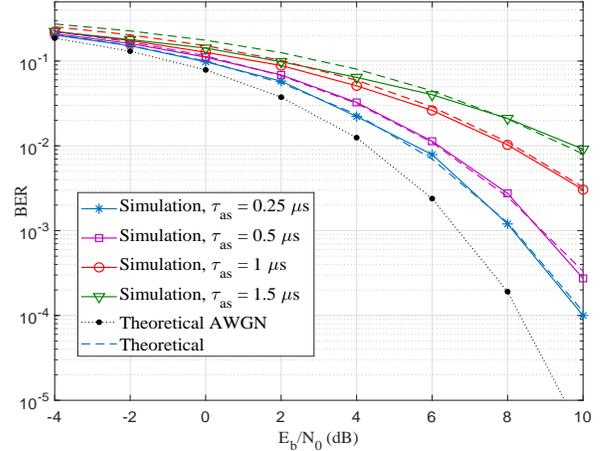}
\caption{BER versus $E_b/N_0$. SIR = 0 dB, $\lambda=2B_s$.}
\label{fig:BERvsSNR_SIR0}
%\vspace{-.25cm}
\end{figure}
%-------------------------------------------------------------------
%-------------------------------------------------------------------
\begin{figure}[t]
\centering
\includegraphics[width=.5\textwidth,height=65mm]{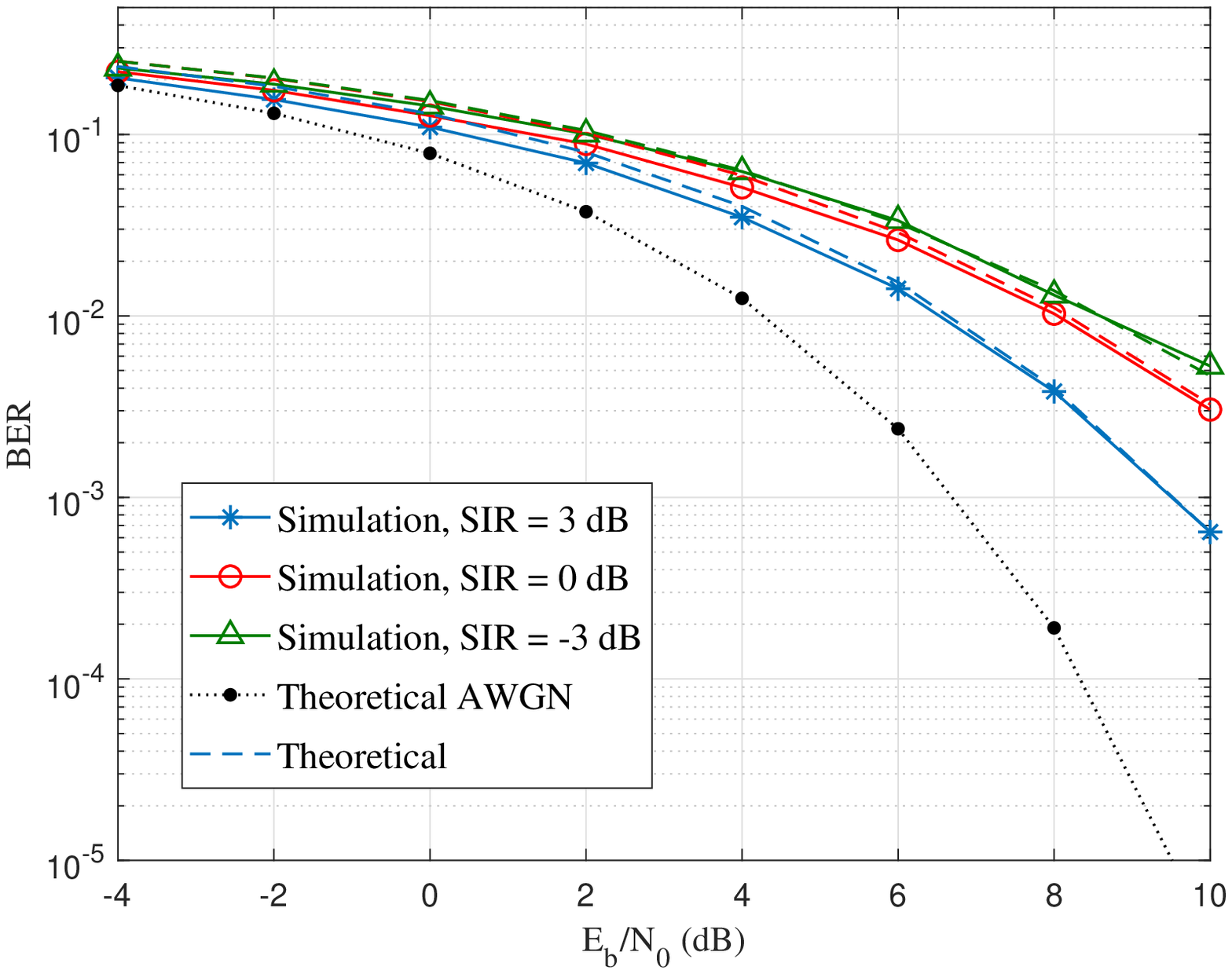}
\caption{BER versus $E_b/N_0$. $\lambda=2B_s$, $\tau_{as}=1\mu {\rm{s}}$.}
\label{fig:BERvsSNR_SIR_303}
%\vspace{-.5cm}
\end{figure}
%-------------------------------------------------------------------

The SNR performance for linear filter, ANDL, blanking, and clipping in various noise compositions is compared in Fig.~\ref{fig:SNR}. According to Fig.~\ref{fig:SNR}, all approaches provide effectively equivalent performance when thermal noise dominates the IN. However, the superiority of the ANDL is highlighted when the IN is dominant and in low SNR (SNR less than zero) its performance is almost insensitive to further increase in the IN power. The potency of the ANDL in IN environment is validated by both simulation and theoretical results. The BER performance of the ANDL in fixed SIR and different duration of IN versus Eb/N0 is shown in Fig.~\ref{fig:BERvsSNR_SIR0}. As expected, we have better performance in short duration IN. Fig.~\ref{fig:BERvsSNR_SIR_303} shows the BER performance of the ANDL in fixed duration of IN and different values of SIR versus Eb/N0. As shown in Fig.~\ref{fig:BERvsSNR_SIR0} and Fig.~\ref{fig:BERvsSNR_SIR_303}, the theoretical results are well aligned with simulation in different scenarios which validate our theoretical calculations.

\begin{figure}%[t]
\centering
\includegraphics[width=.5\textwidth,height=65mm]{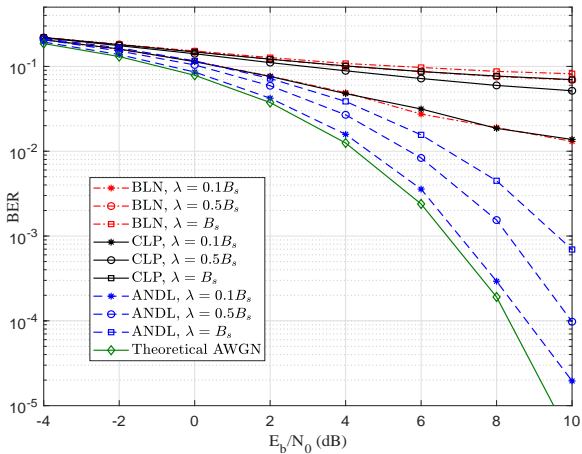}
\caption{BER comparison of ANDL, BLN, and CLP versus $E_b/N_0$ for different values of $\lambda$. SIR = 0 dB, $\tau_{as}=1\mu {\rm{s}}$.}
\label{fig:BER_Comparison_SIR0_Lambda}
%\vspace{-.4cm}
\end{figure}

\begin{figure}%[t]
\centering
\includegraphics[width=.5\textwidth,height=65mm]{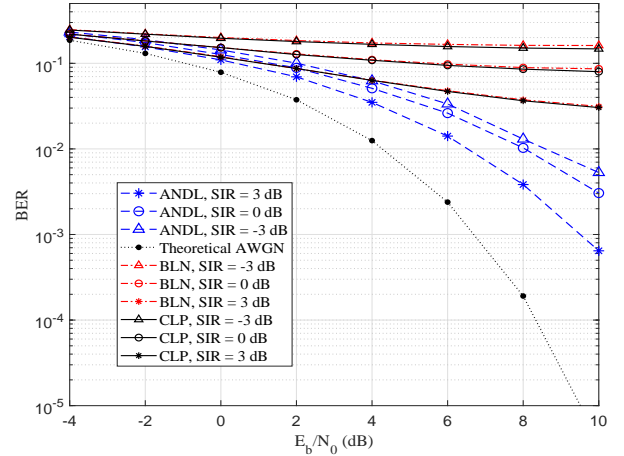}
\caption{BER comparison of ANDL, BLN, and CLP versus $E_b/N_0$ for different values of SIR. $\lambda=2B_s$, $\tau_{as}=1\mu {\rm{s}}$.}
\label{fig:BERvsSNR_fixed_SIR_Comp}
%\vspace{-.4cm}
\end{figure}

Fig.~\ref{fig:BER_Comparison_SIR0_Lambda} compares the BER performance of ANDL with blanking and clipping for different levels of impulsivity ($\lambda$) with $\tau_{as}=1\mu {\rm{s}}$. Fig.~\ref{fig:BER_Comparison_SIR0_Lambda} shows that blanking and clipping are very vulnerable to impulsivity level and their performance is dramatically poor in high impulsive environment. Although, the performance loss of the ANDL with increasing the impulsivity level is also noticeable, still outperforms other approaches in all scenarios. In Fig.~\ref{fig:BERvsSNR_fixed_SIR_Comp}, the BER performance of ANDL for different values of SIR in highly impulsive environments ($\lambda=2B_s$) is compared with blanking and clipping. Fig.~\ref{fig:BERvsSNR_fixed_SIR_Comp} shows that  both blanking and clipping have poor performance and ANDL outperforms them especially at high SNR. The potency of ANDL in reducing the PSD of IN in the signal passband is due to the fact that unlike other nonlinear methods, ANDL is implemented in the analog domain where the outliers are still broadband and distinguishable. Therefore, in highly impulsive environment as shown in Fig.~\ref{fig:BERvsSNR_fixed_SIR_Comp}, ANDL is highly preferable to digital approaches such as blanking and clipping.
%-------------------------------------------------------------------
%-------------------------------------------------------------------
%-------------------------------------------------------------------
%-------------------------------------------------------------------

\section{Conclusion}

In this work, an adaptive analog intermittently nonlinear filter, referred to as Adaptive Nonlinear Differential Limiter (ANDL) is proposed to mitigate impulsive noise (IN) in OFDM-based systems. In addition, an approximation of the ANDL using a piecewise combination of linear filters is used to derive closed-form analytical expressions for the average signal-to-noise ratio (SNR) at the output of the proposed filter. We also show that the theoretical BER results are well aligned with simulation results for different compositions of noise. The theoretical analysis and simulation results show that the ANDL ensures significant improvement in SNR or BER performance in the presence of strong IN component. Moreover, the ANDL outperforms other conventional outlier mitigation methods that exploit amplitude distribution such as blanking and clipping by providing higher output SNR and lower BER in IN environments. It is important to note that the proposed ANDL is totally blind and can be deployed in real-time applications for both sparse and bursty IN scenarios.

%It is important to note that the proposed ANDL can be deployed either as a stand-alone low-cost real-time solution for impulsive noise mitigation, or combined with other interference reduction techniques.

\bibliographystyle{IEEEtran}

\bibliography{IEEEabrv,Reference}

\end{document}